\documentclass[aps, showpacs,pra,twocolumn] {revtex4}
\usepackage{epsf,epsfig}
\usepackage[psamsfonts]{amssymb}
\usepackage{hyperref}
\usepackage{amsmath }
\usepackage{amssymb}
\usepackage{natbib}
\newcommand{\ket}[1]{|#1\rangle}
\newcommand{\bra}[1]{\langle #1|}
\newcommand{\proj}[1]{\ket{#1}\bra{#1}}

\newcommand{\tr}[1]{\mbox{Tr} #1}
\newcommand{\Rtil}{\tilde{\rho}}

\begin{document}
\title{The effect of spin-orbit interaction on entanglement of two-qubit
Heisenberg XYZ systems in an inhomogeneous magnetic field}
\author{Fardin Kheirandish\footnote{fardin$_{-}$kh@phys.ui.ac.ir},
S. Javad Akhtarshenas \footnote{akhtarshenas@phys.ui.ac.ir} and
Hamidreza Mohammadi
\footnote{h.mohammadi@phys.ui.ac.ir}}\affiliation{Department of
Physics, University of Isfahan,
 Hezar Jarib Ave., Isfahan, Iran}

\begin{abstract}
\noindent The role of spin-orbit interaction on the ground state and
thermal entanglement of a Heisenberg XYZ two-qubit system in the
presence of an inhomogeneous magnetic field is investigated. For a
certain value of spin-orbit parameter $D$, the ground state
entanglement tends to vanish suddenly and when $D$ crosses its
critical value $D_c$, the entanglement undergoes a revival. The
maximum value of the entanglement occurs in the revival region. In
the finite temperatures, there are revival regions in $D-T$ plane
where increasing of temperature first increases the entanglement and
then tends to decrease it and ultimately vanishes for temperatures
above a critical temperature. This critical temperature is an
increasing function of $D$, thus the nonzero entanglement can exist
for larger temperatures. In addition, the amount of entanglement in
the revival region depends on the spin-orbit parameter. Also,
entanglement teleportation via the quantum channel constructing by
above system is investigated and influence of spin-orbit interaction
on the fidelity of teleportation and entanglement of replica state
is studied.
\end{abstract}

\maketitle

\section{Introduction}
A. Einstein, B. Podolsky  and N. Rosen, in their famous EPR paradox,
argued that in general two quantum system cannot be separated even
if they are located far from each other \cite{EPR}. E.
Schr\"{o}dinger named this quantum mechanical property as
\textit{Entanglement} \cite{S}. Today, entanglement is a uniquely
quantum mechanical resource that plays a key role in many of the
most interesting applications of quantum computation and quantum
information \cite{NC-book,A-book}. Thus a great deal of efforts have
been devoted to study and characterize the entanglement in the
recent years . The central task of quantum information theory is to
characterize and quantify entanglement of a given system. A mixed
state $\rho$ of a bipartite system is said to be separable or
classically correlated if it can be expressed as a convex
combination of uncorrelated states  $\rho_A$ and $\rho_B$ of each
subsystems i.e. $\rho=\sum\limits_{i} \omega_i \rho_A ^i \otimes
\rho_B ^i$ such that $\omega_i \geq 0$ and $\sum\limits_{i}\omega_i
=1$, otherwise $\rho $ is entangled \cite {CR-2004,A-book}. Many
measures of entanglement have been introduced and analyzed \cite
{NC-book,W,E-thesis}, but the one most relevant to this work is
entanglement of formation, which is intended to quantify the
resources need to create a given entangled state \cite {W}. For the
case of a two-qubit system Wootters has shown that entanglement of
formation can be obtained explicitly as:
\begin{equation}\label{wooters}
E(\rho ) = \Xi [C(\rho )] = h\left( {\frac{{1 + \sqrt {1 + C^2 }
}}{2}} \right),
\end{equation}
where $h(x)=-xlog_2 x - (1-x)log_2 (1-x)$ is the binary entropy
function and
$C(\rho)=\max\{0,2\lambda_{max}-\sum\limits_{i=1}^{4}\lambda_i\}$ is
the concurrence of the state, where $\lambda_i$s are positive square
roots of the eigenvalues of the non-Hermitian matrix $R=\rho \Rtil$,
and $\tilde{\rho}$ is defined by $ \Rtil:=(\sigma^y \otimes
\sigma^y)\rho^* (\sigma^y \otimes \sigma^y)$. The function $\Xi$ is
a monotonically increasing function and ranges from 0 to 1 as C goes
from 0 to 1, so that one can take the concurrence as a measure of
entanglement in its own right. In the case that the state of the
system is pure i.e. $\rho=\proj {\psi}$, $\ket{\psi}=a \ket{00}+b
\ket{01}+c \ket {10}+d \ket{11}$, the above formula is simplified to
$C(\ket{\psi})=2
\mid ad-bc \mid$. \\
The spin chain is the natural candidates for the realization of
entanglement and Heisenberg model is the simplest method for
studying and investigating the behavior of the spin chains . Nielsen
\cite{N} is the  first person who studied the entanglement of
two-qubit Heisenberg XXX- chain modeled with the Hamiltonian $ H= J
\boldsymbol{\sigma _1} \cdot \boldsymbol{\sigma _2}+ \textbf{B}
\cdot (\sigma _1 ^z + \sigma_2 ^z)$. He showed that entanglement in
such systems exists only for antiferromagnatic ($J>0$) case below a
threshold temperature $T_c$. After Nielsen, entanglement in the
two-qubit XXX, XXZ and XY systems in the presence of homogeneous and
inhomogeneous magnetic field has been investigated
\cite{GKV,ABV,KS,SCC,ZL,AK}. The effect of anisotropy due to spin
coupling in the x,y,z direction has also studied in a number of
works \cite{ZSGL,R}. In ref \cite{YGZS} Yang \textit{et al.} have
shown that in XYZ Heisenberg systems, an inhomogeneous external
magnetic field can make the larger revival, improve the critical
temperature and enhance the entanglement. Spin-orbit (SO)
interaction cause another type of anisotropy
\cite{D,M1,M2,M3,BS,BLs,WL}. The effect of SO interaction in the
thermal entanglement of a two-qubit XXX system in the absence of
magnetic field has been studied in \cite{XW}. However, the
entanglement for a XYZ Heisenberg system under an inhomogeneous
magnetic field in the presence of SO interaction has not been
discussed. Therefore, in this  paper we investigate the influence of
SO interaction on the entanglement and entanglement teleportation of
two-qubit system at thermal equilibrium. \\
On the other hand, among the numerous concepts to implement a
quantum bit (qubit), approaches based on semiconductor Quantum Dots
(QDs) offer the great advantage that ultimately a miniaturized
version of quantum computer is feasible. Indeed, at first D. Loss
and D.P. Diviencenzo proposed a quantum computer protocol based on
electron spin trapped in semiconductor QDs in 1998 \cite{LDi,DD,CL}.
Here, the qubit is represented by a single electron in a QD which
can initialize, manipulate and read out by extremely sensitive
devices. Compared with other systems such as quantum optical systems
\cite{X-thesis} and nuclear magnetic resonance (NMR) \cite
{CFH,CGK,JMH}, QDs are argued to be more scalable and has long
decoherence time \cite{NC-book}. The above Heisenberg system is
suitable for modeling  and computing the entanglement of a two-qubit
system represented by two electrons confined in two vertically
coupled quantum  dots (CQDs), respectively. Due to weak lateral
confinement electrons can tunnel from one dot to another dot and
spin-spin and spin-orbit interaction between the two qubits exists.
In GaAs double QDs, hyperfine interactions have been identified to
dominate spin mixing at small magnetic fields, while SO interaction
is not relevant in this regime. However SO interaction and the
coupling magnetic fields are expected to be orders of magnitude
stronger in InAs compared to GaAs \cite{PEL}. SO interaction in such
nanostructures can be
investigated with the help of quantum optical methods \cite{ZZZS}. \\
Taking the advantage of tunability of SO strength
\cite{ZZZS,DE,GBL,FSF}, we show that this type of interaction cause
to enhancement of entanglement in the revival region, increase the
volume of revival region and improve the critical values of other parameters.\\
In the following, as an application of the above system, the
entanglement teleportation of a two-qubit pure state via the above
two-qubit system is investigated and average fidelity between input
and output states is calculated. C. Bennet \textit{et al.} have
shown that two entangled spatially separated particles can be used
for teleportation \cite{BBCJPW}. They also argued that states which
are less entangled still could be used for teleportation but they
reduce "the fidelity of teleportation and/or the range of state
$\ket{\phi}$ which can accurately be teleported". After then S.
Popscu by using hidden variable model, have shown that,
teleportation of a quantum state via a pure classical communication
cannot performed with fidelity larger than $\frac {2}{3}$
\cite{Pop}. Thus, mixed quantum channels which allows to transfer
the quantum information with the fidelity larger than $\frac {2}{3}$
are worthwhile. Horodecki \textit{et al.} have calculated the
optimal fidelity of teleportation for bipartite state acting on $C^d
\otimes C^d$ by using the isomorphism between quantum channels and a
class of bipartite states and twirling operations \cite{HHH}. G.
Bowen and S. Bose have shown that "standard teleportation with an
arbitrary mixed state resource is equivalent to general depolarizing
channel with the probabilities given by the maximally entangled
component of the resource. This enables the usage of any quantum
channel as a generalized depolarizing channel without additional
twirling operation" \cite{BB}. Using the property of linearity of
teleportation process \cite{BBCJPW}, J. Lee and M.S. Kim have shown
that "quantum teleportation preserves the nature of quantum
correlation in the unknown entangled state if the channel is quantum
mechanically correlated" and then they investigate entanglement
teleportation via two copies of werner states \cite{LK}. The
entanglement teleportation via thermally entangled state of a
two-qubit Heisenberg XX-chain and XY-chain has been studied by Ye
Yeo \textit{et al.} \cite{Y,YLLY}. The effect of spin-orbit
interaction on entanglement teleportation on a two-qubit
XXX-Heisenberg chain in the absence of magnetic field is reported by
G. Zhang \cite{Zh}.\\
In this paper we investigate the ability of the above mentioned
two-qubit system for the entanglement teleportation. We show that
spin-orbit interaction and inhomogeneous magnetic field have
effective effect on the entanglement of replica state and the
fidelity of teleportation. A minimal entanglement of the thermal
state in the model is required to realize efficient entanglement
teleportation and we can attain to this minimal entanglement, in the
case of $J_z <0$, by introducing SO interaction.\\
 The paper is organized as follows. In sec. 2 we introduce the Hamiltonian of a two
qubit Heisenberg system under inhomogeneous magnetic field with SO
interaction and write the thermal density matrix of the system
related to this Hamiltonian and ultimately calculate the thermal
concurrence of the system. In subsection  2.1, the ground state
entanglement is calculated and results are plotted in figs. 1-2c. In
subsection 2.2 the finite temperature entanglement of system is
computed. The figs. 3-5 illustrate the obtained results. The
entanglement teleportation of a two-qubit pure state and its
fidelity derived in sec. 3 and results are plotted in figs. 6-8. In
sec. 4 a discussion concludes the paper.

\section{Model and Hamiltonian}

The Hamiltonian of a two-qubit anisotropic Heisenberg XYZ-model in
the presence of inhomogeneous magnetic field and spin-orbit
interaction is:

\begin{eqnarray}\label{Hamiltonian 1}
 H &=& {\textstyle{1 \over 2}}(J_x \,\sigma _1^x \sigma _2^x \,
 + J_y \,\sigma _1^y \sigma _2^y  + J_z \,\sigma _1^z \sigma _2^z
 +\textbf {B} _1 \cdot \boldsymbol {\sigma} _1
  + \textbf{B} _2 \cdot \boldsymbol {\sigma} _2
 \nonumber\\&+& \textbf{D} \cdot (\boldsymbol {\sigma} _1  \times
  \boldsymbol{\sigma}  _2
 )+ \delta \,\, \boldsymbol {\sigma}_1 \cdot \overline{\mathbf{\Gamma}}\cdot \boldsymbol
 {\sigma}_2),
 \end{eqnarray}

where $\boldsymbol{\sigma}_{j}=(\sigma^{x}_{j}, \sigma^{y}_{j},
\sigma^{z}_{j})$ is the vector of Pauli matrices, $\textbf{B}_j
\,(j=1,2)$  is the magnetic field on site j, $J_\mu \,(\mu=x,y,z)$
are the real coupling coefficients (the chain is anti-ferromagnetic
(AFM) for $J_\mu >0$ and ferromagnetic (FM) for $J_\mu <0$)  and
$\textbf{D}$ is called Dzyaloshinski-Moriya vector, which is first
order in spin-orbit coupling and is proportional to the coupling
coefficients ($J_\mu$) and $\overline{\mathbf{\Gamma}}$ is symmetric
tensor which is second order in spin-orbit coupling
\cite{D,M1,M2,M3}. For simplicity we assume $\textbf{B}_j =B_j \,
\boldsymbol{\hat{z}}$ such that $B_1 =B+b$ and $B_2 =B-b$, where b
indicates the amount of inhomogeneity of magnetic field. The vector
$\textbf{D}$ and the parameter $\delta$ are dimensionless, in system
like coupled GaAs quantum dots
$\boldsymbol{|}\textbf{D}\boldsymbol{|}$ is of order of a few
percent, while the order of last term is $10^{-4}$ and is
negligible. If $\textbf{D}= J_z D \, \boldsymbol{\hat{z}}$ and
ignore the second order spin-orbit coupling, then the above
Hamiltonian can be expressed as:

\begin{eqnarray}\label{Hamiltonian 2}
 H &=& J\gamma (\sigma _1^ +  \sigma _2^ +   + \sigma_1^ -  \sigma _2^ -  )
 + (J + iJ_z D)\sigma _1^ +\sigma _2^ -   + (J - iJ_z D)\sigma _1^ -  \sigma _2^ +
  \nonumber\\&+& \frac{{J_z }}{2}\sigma _1^z \sigma _2^z  + (\frac{{B + b}}{2})\sigma _1^z
  + (\frac{{B - b}}{2})\sigma _2^z,
 \end{eqnarray}

where $J :=\frac{J_x + J_y}{2}$, is the mean coupling coefficient in
the XY-plane, $\gamma :=\frac{J_x - J_y}{J_x + J_y}$, specifies the
amount of anisotropy in the XY-plane (partial anisotropy, $-1\leq
\gamma \leq 1$) and $\sigma^\pm=\frac{1}{2}(\sigma^x \pm \sigma^y)$
are lowering and raising
operators.\\
The Hamiltonian (\ref{Hamiltonian 2}), in the standard basis
$\{\ket{00},\ket{01},\ket{10},\ket{11}\}$, has the following matrix
form:

\begin{eqnarray}\label{Hamiltonian 3}
H = \left( {\begin{array}{*{20}c}
   {\frac{{1 }}{2} J_z + B} & 0 & 0 & {J\gamma }  \\
   0 & {- \frac{{1 }}{2} J_z + b} & {J + iJ_z D} & 0  \\
   0 & {J - iJ_z D} & {- \frac{{1 }}{2} J_z - b} & 0  \\
   {J\gamma } & 0 & 0 & {\frac{{1 }}{2} J_z - B}  \\
\end{array}} \right).
\end{eqnarray}
The spectum of H is easily obtained as
\begin{eqnarray}\label{spectrum}
\,H \ket{\psi ^ \pm}  = \varepsilon _{1,2} \ket{\psi ^ \pm }\,, \nonumber \\
\\ \nonumber H \ket{\Sigma^ \pm }  = \varepsilon _{3,4} \ket{\Sigma ^
\pm}\,,
\end{eqnarray}

where the eigenstates and the corresponding eigenvalues are,
respectively
\begin{eqnarray}\label{eigenstates}
\begin{array}{l}
 \ket{\psi ^ \pm}  = N^ \pm  ( - (\frac{{b \pm \xi }}{{J - iJ_z D}})\ket{01}  + \ket{10})\,,
 \\ \\
 \ket{\Sigma ^ \pm}  = M^ \pm  ( - (\frac{{B \pm \eta }}{{J\gamma }})\ket{00}  + \ket{11} )\,, \\
 \end{array}
\end{eqnarray}
and
\begin{eqnarray}\label{eigenvalues}
\begin{array}{l}
 \varepsilon _{1,2}  =  - \frac{{1}}{2} J_z \pm \xi\,,  \\ \\
 \varepsilon _{3,4}  = \frac{{1}}{2} J_z \pm \eta\,.  \\
 \end{array}
\end{eqnarray}
In the above equations $N^\pm =\frac{1}{\sqrt{1+\frac{{(b \pm
\xi)^2}}{J^2 +(J_z D)^2}}}$ and $M^\pm =\frac{1}{\sqrt{1+(\frac{{B
\pm \eta}}{J \gamma})^2}}$ are the normalization constants. Here we
define, $ \xi  := \sqrt {b^2  + J^2  + (J_z D)^2 }$ and
$\eta := \sqrt {B^2  + (J\gamma )^2 }$, for later convenience.\\
The state (density matrix) of a system in equilibrium at temperature
T is $\rho = Z^{-1} \exp (-\frac {H}{k_B T})$, where Z is the
partition function of the system and $k_B$ is the Boltzman constant.
For simplicity we take $k_B =1$. In the standard basis, the density
matrix of the system in the thermal equilibrium can be written as:
\begin{eqnarray}\label{density matrix}
\rho_T  = \left( {\begin{array}{*{20}c}
   {\mu _ +  } & 0 & 0 & \nu   \\
   0 & {w_1 } & z & 0  \\
   0 & {z^* } & {w_2 } & 0  \\
   \nu  & 0 & 0 & {\mu _ -  }  \\
\end{array}} \right),
\end{eqnarray}
where
\begin{eqnarray}\label{DM component}
\begin{array}{l}
 \mu _ \pm   = \frac{{e^{ - {\textstyle{{\beta J_z } \over 2}}} }}{Z}
 (\cosh \beta \eta  \mp \frac{B}{\eta }\sinh \beta \eta )\,, \\
 \\w_{1,2}  = \frac{{e^{{\textstyle{{\beta J_z } \over 2}}} }}{Z}
 (\cosh \beta \xi  \mp \frac{b}{\xi }\sinh \beta \xi )\,, \\
 \\\nu  =  - \frac{{J\gamma \,e^{ - {\textstyle{{\beta J_z }
 \over 2}}} }}{{Z\eta }}\sinh \beta \eta\,,  \\
 \\z =  - \frac{{(J + iJ_z D)\,e^{ {\textstyle{{\beta J_z }
  \over 2}}} }}{{Z\xi }}\sinh \beta \xi\,,  \\
 \end{array}
\end{eqnarray}
and $Z= 2 e^{\frac{\beta J_z}{2}}(\cosh{\beta \xi}+e^{- \beta
J_z} \cosh{\beta \eta})$ .\\
In what follows, our purpose is to quantify the amount of
entanglement of the above two-qubit system versus the parameters of
the system, with the main concerning on $D$ . For density matrix in
the form (\ref{density matrix}), one can show, by straight forward
calculations, that the square roots of the eigenvalues of matrix
$R=\rho \Rtil$ are:

\begin{widetext}
\begin{eqnarray} \label{lambda}
 \lambda _{1,2}&  =& \mid \sqrt {w_1 w_2 } \pm \mid z\mid \mid
\nonumber \\&=& \frac{{e^{{\textstyle{{\beta J_z } \over 2}}} }}{\xi
 Z}\left| {\left.
  {\sqrt {\xi ^2 +  J^2  + (J_z D)^2  \sinh ^2
  \beta \xi }  \pm  \sqrt {J^2  + (J_z D)^2 } \sinh
  \beta \xi } \right|} \right.,
\nonumber \\ \\
\lambda_{3,4}& =& \mid \sqrt {\mu _ +  \mu _ -  }  \mp
\nu\mid\nonumber \\ \nonumber &\, =& \frac{{e^{ - {\textstyle{{\beta
J_z } \over 2}}} }}{\eta Z}\left|
  {\left. {\sqrt {\eta^2 +  (J\gamma)^2 \sinh ^2 \beta \eta }
    \mp J\gamma \sinh \beta \eta } \right|} \right..
\end{eqnarray}
\end{widetext}

\begin{figure}
\epsfxsize=6cm \ \centerline{\hspace{0cm}\epsfbox{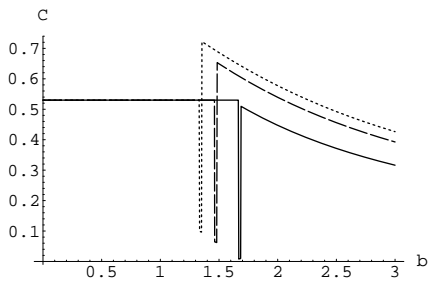}} \
\caption{The ground state concurrence vs. $b$ for $D=0$ (solid
line), $D=0.8$ (dashed line) and $D=1$ (doted line) where $B=0.8,
J=1, J_z=-1$ and $\gamma=0.5$.}\label{f1}
\end{figure}

\begin{figure}
\epsfxsize=6cm \ \centerline{\hspace{0cm}\epsfbox{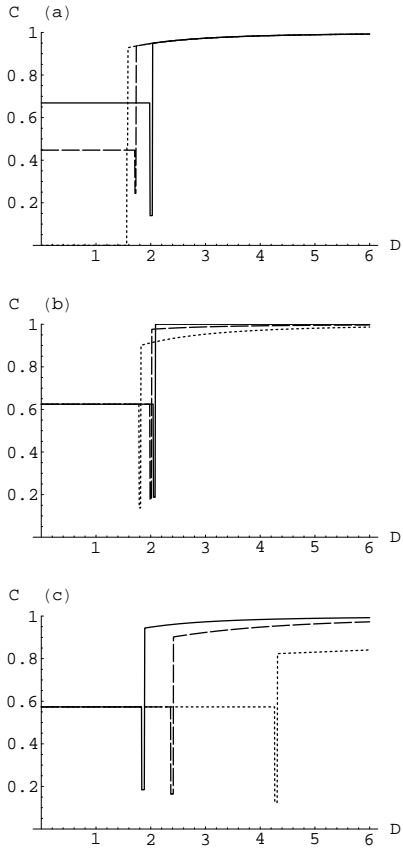}} \
\caption{The concurrence of the ground state vs. $D$. a) for $b=0$
(solid line), $b=0.5$ (dashed line) and $b=1$ (doted line), $J_z=-1$
and $\gamma=0.8$  b) for $\gamma=0.9$ (solid line), $\gamma=0.5$
(dashed line) and $\gamma=0$ (doted line),  $J_z=-1$ and $b=0.75$ c)
for $J_z=-1$ (solid line), $J_z=0.5$ (dashed line) and $J_z=0.1$
(doted line), $b=0.75$ and $\gamma=0.7$.}\label{f2}
\end{figure}

 Now, it is easy to calculate the concurrence. Without loss of generality we can assume
 $J>0$ and  $\gamma >0$, since the above formula are invariant under substitution $J\rightarrow-J$
  and $\gamma \rightarrow -\gamma$.
For the special case $D=0$, these equations give the same results as
ref \cite{YGZS}. The obtained results are given in the following
subsections.
\subsection{Ground state entanglement and critical parameters}
The behavior of the system  at the quantum phase transition (QPT)
point \footnote {Hint: A phase transition can, strictly speaking,
only occur in the thermodynamic limit of $\sim 10^{23}$ particles.
However, one may see traces of what may become a phase transition at
smaller systems, it is important to emphasize that no actual phase
transition can occur in this system \cite {S-thesis, S-book}.}, such
as $B=B_c,\,b=b_c\,$ or $\, D=D_c$, may be determined from the
density matrix at $T=0$, at which the system is in its ground state.
If we consider $T\rightarrow 0$ (or $\beta \rightarrow \infty$) then
the concurrence C can be written analytically as:

\begin{widetext}
\begin{eqnarray} \label{C(T=0)}
C(T = 0) = \left\{ \begin{array}{l}
 \left| {\left. {\frac{{J\gamma }}{\eta }} \right|} \right.\,\,\,\,\,\,\,\,\,\,\,\,\,\,\,\,\,\,\,\,\,\,\,\,\,\,\,\,\,\,\,\,\,\,\,\,\,\,\,\,\,\,\,\,\,\,\,\,\,\,\,\,\,\,\,\,\,\,\,\,\,\,\,\,\,\,\,\,\,if\,\,\xi  < \eta  - J_z  \\
 {\textstyle{1 \over 2}}\left| {\left. {|\frac{{J\gamma }}{\eta }\,| - \frac{{\sqrt {J^2  + (J_z D)^2 } }}{\xi }} \right|\,\,\,\,\,\,\,\,\,\,\,\,\,\,\,\,\,\,\,\,\,\,\,\,\,\,\,if\,\,\xi  = \eta  - J_z } \right. \\
 \left| {\left. {\frac{{\sqrt {J^2  + (J_z D)^2 } }}{\xi }} \right|} \right.\,\,\,\,\,\,\,\,\,\,\,\,\,\,\,\,\,\,\,\,\,\,\,\,\,\,\,\,\,\,\,\,\,\,\,\,\,\,\,\,\,\,\,\,\,\,\,if\,\,\xi  > \eta  - J_z  \\
 \end{array} \right.\,\,\,\,\,\,\,\,\,\,\,\,\,\,\,\,\,\,\,
\end{eqnarray}
\end{widetext}

 Also, this formula can be derived by calculating entanglement of ground
state of H directly: When $\epsilon_4 < \epsilon_2$ (or $\xi < \eta
-J_z$) the ground state of H is $\ket{\Sigma^-} $ and then $C=\mid
\frac {J \gamma}{\eta} \mid $ and when $\epsilon_2 < \epsilon_4$ (or
$\xi > \eta -J_z$ ) the ground state of H is $\ket{\psi^-} $ and
then $C=\mid \frac {\sqrt{J^2 +(J_z D)^2}}{\xi} \mid $. On the other
hand, at critical point (where $\epsilon_4 = \epsilon_2$ or $\xi <
\eta -J_z$), the ground state of the system is an equally mixture of
$\ket{\Sigma^-} $ and $\ket{\psi^-} $ i.e.
 $\ket{GS} =(\ket{\Sigma ^-} +\ket{\psi ^-})/2$ and then $C = {\textstyle{1 \over 2}}\left| {\left.
  {\left| {\left. {\frac{{J\gamma }}{\eta }}\right|} \right.- \frac{{\sqrt {J^2
  + (J_z D)^2 } }}{\xi }} \right|} \right.$.\\
The concurrence $C$ as a function of $b$ (at $T=0$ ) for three
values of $D\,( D=0, 0.8, 1$ ) are plotted in fig. 1. With
increasing $b$, the concurrence $C$ is initially constant and equal
to $C=\mid \frac {J \gamma}{\eta} \mid = 0.53$, then drops suddenly
at critical value of b $(b_c =\sqrt{(\eta - J_z)^2 -((J_z
D)^2+J^2)})$. At this point ($T=0, b=b_c$ ), the concurrence becomes
a non-analytical function of $b$ and QPT occurs \cite {YGZS}. For
$b>b_c$, concurrence $C$ undergoes a revival before decreasing to
zero. The amount of concurrence at revival region depends on $D$, by
increasing $D$ the revival is greater. Furthermore, by increasing
$D$, $b_c$ decreases i.e. for larger $D$, the critical point and
hence revival phenomenon occurs in smaller $b$. The role of $D$ is
more obvious in figs. 2a-2c where concurrence is plotted in terms of
$D$. These figures show that, when $D$ reaches its critical value
defined by
\begin{eqnarray}\label{Dc}
D_c=\sqrt{(\eta - J_z)^2 -(b^2+J^2))}/|J_z|,
\end{eqnarray}
the concurrence drops and exhibit a revival phenomenon when $D$
crosses its critical value $D_c$. In the revival region (larger $D$)
the concurrence reaches its maximum value ($C=1$). Fig. 2a
illustrates the ground state concurrence variation versus $D$ for
three values of b, by decreasing $b$, the critical value of $D$
($D_c$ ) increases. Fig. 2b shows ground state concurrence versus
$D$ for three values of $\gamma$, by increasing $\gamma$, $D_c$
increases (amount of $\gamma$ determines the value of entanglement
before reaching the critical point). In fig. 2c the ground state
concurrence versus $D$ is plotted for three values of $J_z$, when
$\mid J_z \mid $ increases, $D_c$ also increases.\\
Hence, we can control value of $D_c$ by adjusting the parameters of
the system such as $b, \gamma$ and $J_z$. We demand to decrease
$D_c$ (because for $D>D_c$ concurrence will be maximized). Therefore
we should choose, $\mid J_z \mid =1, \gamma$ small and $b$ as large
as possible.

\subsection{Thermal entanglement}
 Since the relative magnitude of $\lambda_i (i=1, 2, 3, 4)$ depends on the parameters involved,
they cannot be ordered by magnitude without knowing the values of
the parameters. This prevents one from writing an analytical
expression for concurrence. For particular parameters, $C$ can be
evaluated numerically. The role of each parameter can be seen by
fixing other parameters and drawing variation of entanglement for
specific values of that parameter. Cross influence of two parameters
can be shown in 3D plots of entanglement. Thermal entanglement
versus system's parameters is depicted in figures 3-5. The
concurrence as a function of $T$ and $D$ is shown in fig. 3. The
analysis of results of this figure are given in the following. Let $
C_j'  = 2\lambda _j  - \sum\limits_{i = 1}^4 {\lambda _i }$ ($j =1,
2, 3, 4$), it is evident that the function $C_j=max\{0,C_j'\}$ is
the concurrence of the system if and only if $\lambda _j  = \lambda
_{\max }  = \max \{ \lambda _1, \lambda _4 \}$ (Since $ J, \gamma
>0$ we have $\lambda_4>\lambda_3$ and $\lambda_1>\lambda_2$). We can
divide the regions of fig. 3 in four parts:\\

\begin{figure}
\epsfxsize=10cm \ \centerline{\hspace{0cm}\epsfbox{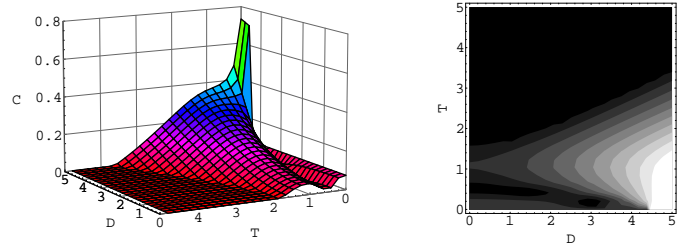}} \
\caption{color online) Thermal concurrence vs. T and D. Where $B=4,
b=2.5, J=1$ and $\gamma=0.3$.}\label{f3}
\end{figure}

\begin{figure}
\epsfxsize=10cm \ \centerline{\hspace{0cm}\epsfbox{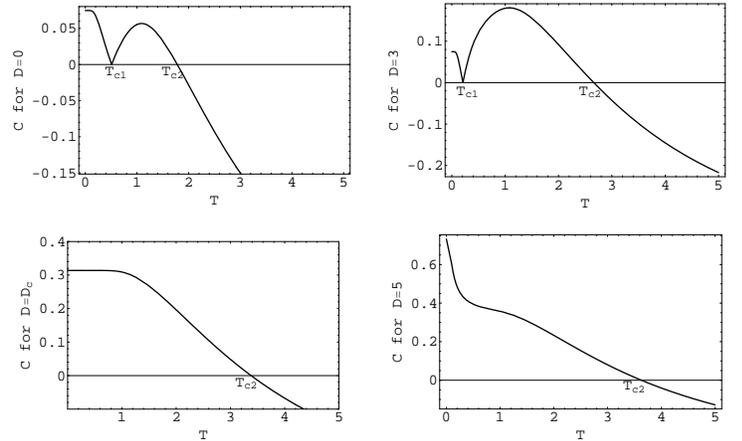}} \
\caption{Thermal concurrence vs. $T$ for different values of $D$.
The parameters are the same as fig. \ref{f3}. In this case
$D_c\simeq4.51$.}\label{f4}
\end{figure}

 \textit{i) Region for which $T<T_{c1}$(see below) and
$D<D_{c}$}; in this region $\lambda_{max}=\lambda_4$ and then $C_4'
= \lambda _4  - \lambda _3  - \lambda _2  - \lambda _1$ determines
the amount of entanglement. According to equation (\ref{lambda}),
the values of $\lambda_4$ depends on $\gamma$ and $\eta$ (or
equivalently $\gamma$ and $B$ ). Thus in this region we can
manipulate the amount of entanglement by adjusting $\gamma$ and $B$.
The first critical temperature ($T_{c1}$ ), is the point at which
$\lambda _{\max }  = \lambda _4 =
\lambda _1$ and hence $C=0$.\\
 \textit{ii) Revival region, for which $T_{c1}<T<T_{c2}$ (see below) and $D<D_{c}$};
in this region $\lambda_{max}=\lambda_1$ and hence $C_1'$ determines
the amount of entanglement. According to equation (\ref{lambda}),
the value of $\lambda _1$  is adjustable by changing values of $D$
and $\xi$ (or equivalently $D$ and $b$) and hence the parameters $D$
and $b$ play an important role in quantifying the amount of
entanglement, in this region. For $T>T_{c1}$, the value of
$\lambda_1$ and also the rate of enhancement of the function
$\lambda_1$ with $T$ increases as $D$ increases. Enhancement of the
function $\lambda_1$ with $T$, cause  $C_1'$ rise to a positive
number and thus the entanglement undergoes a revival. Since the rate
of enhancement of the function $\lambda_1$ with $T$ is an increasing
function of $D$, the amount of revival increases as $D$ increases.
When $T$ reaches the value $T=T_{c2}$ (second critical temperature),
$\lambda_1$ tend to zero again and thus the entanglement vanishes.\\
 \textit{iii) Region for which $D>D_c$, for all values of $T$}; in this region
 $\lambda_{max}=\lambda_1$ and $C_1'$ is the entanglement indicator.
 The maximum value of the entanglement occurs in this
region. At zero temperature the entanglement has its maximum value
and by increasing the temperature the system loses its entanglement
and ultimately vanishes at $T=T_{c2}$. In this region no revival phenomenon occurs.\\
\textit{iv) Region for which $T\geq T_{c2}$ for all values of $D$};
in this region all values of $C_j'$ ($j =1, 2, 3, 4$) have negative
values and then the entanglement is zero for all values of $D$ and the other parameters. \\
Notice that, $T_{c1}$ and $T_{c2}$  are sensitive functions of $D$.
In fig. 4, we try to demonstrate these facts, by illustrating the
function of thermal concurrence vs. $T$ for few values of $D$. This
figure shows that for $D<D_c$, $T_{c1}$ is a decreasing function of
$D$ but for $D>D_c$, $T_{c1}$ is undefined. In contrast, $T_{c2}$ is
an increasing function of $D$ for all values of $T$, i.e. we can
create and maintain the nonzero entanglement at larger temperatures.
In summary, fig. 3 shows that:
 For $T_{c1}<T<T_{c2}$ and $D<D_c$, there are regions in $D-T$ plane
where increasing of temperature first increases the entanglement
(revival region) and then tends to decrease the entanglement and
ultimately for $T>T_{c2}$ entanglement vanishes. The maximum
entanglement exists at zero temperature and for large $D$. In the
revival region and region for which $D>D_c$ , increasing $D$ causes
the entanglement to increase. In this region,
 $T_{c1}$ decreases and $T_{c2}$ increases as $D$ increases and
hence the width of the revival region increases. In all regions
$T_{c2}$ is an increasing function of $D$, thus when $D$ is large
enough, the entanglement can exists for larger temperatures.
Furthermore, the parameter $D$ plays the role of parameter b.
 Fig. 5a, shows the variation of entanglement as a function of $b$ (inhomogeneity
of magnetic field) and $D$. For fixed $D$, there are three region in
this figure \textit{i) Main region where $b<b_c$}; in this region
entanglement is constant \textit{ii) Collapse region where $b=b_c$};
in this region entanglement decreases suddenly \textit{iii) Revival
region where $b>b_c$}; in this region entanglement undergoes a
revival. Amount of $D$ determines $b_c$ and hence edge of revival
region. Furthermore, in the revival region increasing $D$ increases
the entanglement. In fig. 5b, thermal entanglement is plotted vs.
magnetic field ($B$) and $D$. The role of $D$ is similar to its role
in fig. 5a, enhancement of $D$ improves $B_c$ and increases amount
of entanglement in the revival region.\\
One can use the above entangled two-qubit system for performance the
teleportation protocols. The next section is spend to this subject.

\begin{figure}
\epsfxsize=10cm \ \centerline{\hspace{0cm}\epsfbox{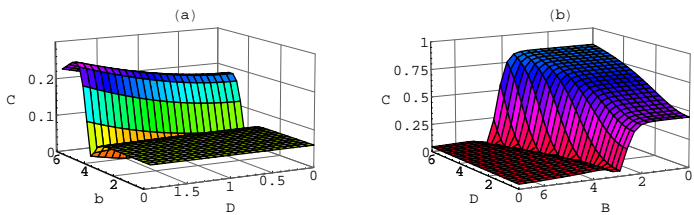}} \
\caption{(color online) (a) Thermal concurrence vs. b and D, where
B=5, (b) Thermal concurrence vs. B and D, where b=2. Here $ T=0.3,
J=1, J_z=0.5$ and $\gamma=0.3$ }\label{f5}
\end{figure}

\section{Thermal Entanglement teleportation}
For the entanglement teleportation of a whole two-qubit system, a
thermal mixed state in Heisenberg spin chain can be considered as a
general depolarizing channel. Now we consider Lee and Kim's two
qubit teleportation protocol ($\textit{P}_1$), and use two copies of
the above two-qubit thermal state, $\rho_T \otimes \rho_T$, as
resource \cite {LK}. Similar to standard teleportation, entanglement
teleportation for the mixed channel of an input entangled state is
destroyed and its replica state appears at the remote place after
applying a local measurement in the form of linear operators. We
consider as input a two-qubit in the special pure state $\ket {
\psi_{in}} = \cos \theta/2  \ket{10} + e^{i \phi} \sin \theta/2
\ket{01}\,(0\leq \theta \leq \pi , 0\leq \phi \leq 2 \pi)$. The
density matrix related to $\ket{\psi_{in}} $ is in the form:
\begin{eqnarray} \label{input}
\rho _{in}  = \left( {\begin{array}{*{20}c}
   0 & 0 & 0 & 0  \\
   0 & a & c & 0  \\
   0 & {c^* } & b & 0  \\
   0 & 0 & 0 & 0  \\
\end{array}} \right),
\end{eqnarray}
where $a=\sin ^2 \theta /2, b=\cos ^2 \theta /2$ and $c=\frac{1} {2}
e^{-i \phi} \sin \theta$. Therefore concurrence of initial state is
$C_{in}=2 \mid e^{i \phi} \sin \theta /2 \cos \theta /2 \mid=\sin
\theta$. The output (replica) state $\rho_{out}$ can be obtained by
applying joint measurement and local unitary transformation on input
state $\rho_{in}$. Thus the out put state is given by \cite {BB}
\begin{eqnarray} \label{output1}
\rho _{out}  = \sum\limits_{\mu,\nu}^{} {p_{\mu \nu} (\sigma _{\mu}
\otimes \sigma _{\nu} )} \rho _{in} (\sigma _{\mu}  \otimes \sigma
_{\nu}),
\end{eqnarray}
where $ \mu, \nu =0,x,y,z$ ($\sigma^0=I$), $p_{\mu \nu}= tr [
E^{\mu} \rho_{channel}] tr[E^{\nu} \rho_{channel}]$ such that
$\sum\limits_{\nu,\mu}^{} {p_{\mu \nu}}=1$ and $\rho_{channel}$
represent the state of channel which used for teleportation. Here
$E^0=\proj{\Psi^-}$, $E^1=\proj{\Phi^-}$, $E^0=\proj{\Phi^+}$ and
$E^0=\proj{\Psi^+}$ where $\ket{\Psi^\pm}=\frac{(\ket{01} \pm
\ket{10})} {\sqrt{2}}$ and $\ket{\Phi^\pm}=\frac{(\ket{00} \pm
\ket{11})} {\sqrt{2}}$
are the Bell states.\\
By considering  the two-qubit spin system as a quantum channel, the
state of channel is $ \rho _{channel} = \rho _T$ given in the
equation (\ref{density matrix}) and hence one can obtain
$\rho_{out}$ as
\begin{eqnarray} \label{output2}
\rho _{out}  = \left( {\begin{array}{*{20}c}
   \alpha  & 0 & 0 & \kappa   \\
   0 & {a'} & {c'} & 0  \\
   0 & {c'^* } & {b'} & 0  \\
   \kappa  & 0 & 0 & \alpha   \\
\end{array}} \right),
\end{eqnarray}
where
\begin{eqnarray} \label{output2 components}
\begin{array}{l}
 \alpha  = (w_1  + w_2 )(\mu ^ +   + \mu ^ -  ), \\
\\
 \kappa  = 4\,{\mathop{\rm Re}\nolimits} [z]\,\nu \,\cos \phi \,\sin \theta,\\
 \\
 a' = (\mu ^ +   + \mu ^ -  )^2 \cos ^2 {\textstyle{\theta  \over 2}} + (w_1  + w_2 )^2 \,\sin ^2 {\textstyle{\theta  \over 2}}, \\
 \\
 b' = (w_1  + w_2 )^2 \,\cos ^2 {\textstyle{\theta  \over 2}} + (\mu ^ +   + \mu ^ -  )^2 \sin ^2 {\textstyle{\theta  \over 2}}, \\
 \\
 c' = 2\,e^{ - i\phi } (({\mathop{\rm Re}\nolimits} [z])^2  + e^{2i\phi } \nu ^2 )\,\sin \theta. \\
 \end{array}
\end{eqnarray}
Following the above we can determine concurrence of out put state by
calculating positive square roots of $R_{out}=\rho_{out} \tilde
\rho_{out}$, i.e. $\lambda_i'$s. It is easy to show

\begin{widetext}
\begin{eqnarray} \label{lambda'}
 \!\! \lambda _{1,2}'  &=& \left| {\frac{{\sqrt {C_{in}^2 (\cosh ^2 \beta \eta  - e^{2\beta J_z }
 \cosh ^2 \beta \xi )^2  + 4e^{2\beta J_z } \cosh ^2 \beta \eta \cosh ^2 \beta \xi } }}
 {{2(\cosh \beta \eta  + e^{\beta J_z } \cosh \beta \xi )^2 }}}
 \right. \nonumber \\
&&\pm \left. {  \frac{{|C_{in} (({\textstyle{{J\gamma } \over \eta
}})^2 e^{2 i\phi} \sinh ^2 \beta \eta  + ({\textstyle{J \over \xi
}})^2 e^{2\beta J_z } \sinh ^2 \beta \xi  )|}}{{2(\cosh \beta \eta +
e^{\beta J_z } \cosh \beta \xi )^2 }}}
  \right|, \\ \nonumber
   \lambda _{3,4}' &=& \left| {\frac{{\cosh \beta \eta \,\cosh \beta \xi  \pm C_{in}
 ({\textstyle{{J\gamma } \over \eta }})({\textstyle{J \over \xi }})\sinh \beta \eta
  \sinh \beta \xi \cos \phi }}{{(\cosh \beta \eta  + e^{\beta J_z } \cosh \beta \xi )^2 }}}
  \right|e^{\beta J_z }.
\end{eqnarray}
\end{widetext}

Thus $C_{out}  = C(\rho _{out} ) = \max \{ 0,2\lambda _{\max }'
-\sum\limits_{i = 1}^4 {\lambda _i' } \} $ is computable when the
parameters of channel are known. The function $C_{out}$ is dependent
on the entanglement of initial state and the parameters of the
channel (which determine the entanglement of channel). The $C_{out}$
is nonzero only for particular choice of channel's parameters for
which $C_{channel}$ is greater than a critical value. The figs. 6
and 8 depict behavior of $C_{out}$ versus the parameters of the
channel and $C_{in}$. For the case $J_z >0$, the entanglement of
replica state ($C_{out}$) increases linearly as $C_{in}$ increases.
The rate of this enhancement is determined by $D$ (indeed, by
$C_{channel}(D)$). But for the case $J_z<0$, $C_{out}$ is zero for
small values of $D$. As $D$ crosses  a threshold value, $C_{out}$
increase when $C_{in}$ increase with the rate determined by amount
of $D$ (equivalently $C_{channel}(D)$). The figs. 6b and 6c show
that the parameter of inhomogeneity ($b$) can plays role of $D$.
\\

\begin{figure}
\epsfxsize=10cm \ \centerline{\hspace{0cm}\epsfbox{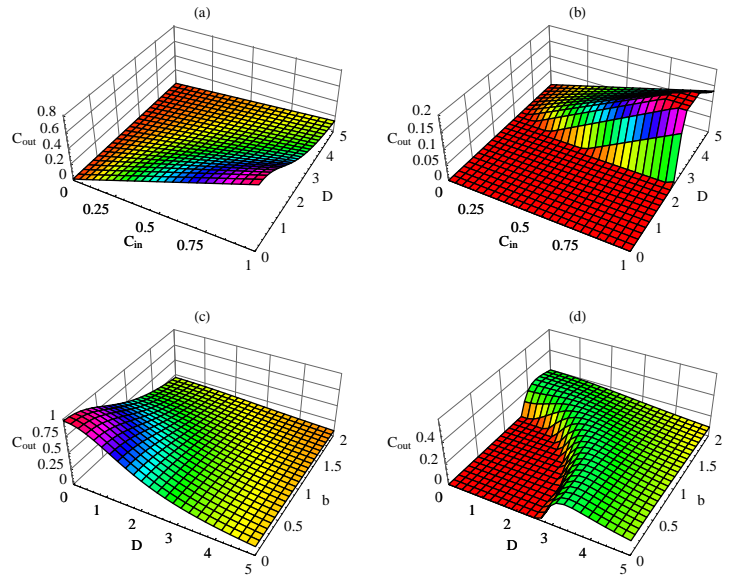}} \
\caption{(color online) Entanglement of output state ($C_{out}$) vs.
the channel's parameters and $C_{in}$. $C_{out}$ vs. $D$ and
$C_{in}$ for a) $b=0.5$ and $J_z >0$ and b)  $b=0.5$ and $J_z <0$.
  Also, $C_{out}$ vs. $b$ and $D$ for c) $C_{in}=1$ and $J_z >0$ and
d)$C_{in}=1$ and $J_z <0$. Where $J=1, T=0.1, B=1$ and
$\gamma=0.3$.}\label{f6}
\end{figure}

\textit{-The Fidelity of entanglement teleportation:} Fidelity
between $\rho_{in}$ and $\rho_{out}$ characterizes the quality of
teleported state $\rho_{out}$. When the input state is a pure state,
we can apply the concept of fidelity as a useful indicator of
teleportation performance of a quantum channel \cite {Y,RJ}. The
maximum fidelity of
 $\rho_{in}$ and  $\rho_{out}$ is defined to be
\begin{eqnarray} \label{fidelity1}
F(\rho _{in} ,\rho _{out} ) &=& \{ \tr{[\sqrt {(\rho _{in}
)^{{\textstyle{1 \over 2}}} \rho _{out} (\rho _{in} )^{{\textstyle{1
\over 2}}} } ]\}} ^2 \nonumber \\ &=& \bra{\psi_{in}}
\rho_{out}\ket{\psi_{in}}.
\end{eqnarray}
By substituting $\rho_{in}$ and $\rho_{out}$ from above, we have

\begin{eqnarray} \label{fidelity2}
F(\rho _{in} ,\rho _{out} ) = a'\,\sin ^2 {\textstyle{\theta  \over
2}} + b'\,\cos ^2 {\textstyle{\theta  \over 2}} + {\mathop{\rm
Re}\nolimits} [c'e^{ - i\phi } ]\,\sin \theta,
\end{eqnarray}

simplifying the above formula we find that the maximum fidelity
$F(\rho_{in},\rho_{out})$ depends on initial entanglement($C_{in}$):
\begin{eqnarray} \label{fidelity3}
F(\rho _{in} ,\rho _{out} ) ={\textit{f}}^{\,\,c} +{\textit{f}}
^{\,\,q}\,\, C_{in}^2\,,
\end{eqnarray}
where ${\textit{f}}^{\,\,c}=(w_1+w_2)^2$ and ${\textit{f}}
^{\,\,q}=\frac{1}{2}-(w_1+w_2)+2 (\nu ^2 \cos 2 \phi +(Re[z])^2)$.
 The functions ${\textit{f}}^{\,\,c}$  and ${\textit{f}}^{\,\,q}$ are
 dependent on the channel's parameters only (we consider $\phi=0$) and hence are relate to
the entanglement of the channel. This formula is the same as the
results of ref. \cite{LK}\footnote {In the ref. \cite{LK}, Lee and
Kim use the negativity as a measure of entanglement and the werner
states as resource. In $2\times2$ systems, the negativity coincides
with concurrence for pure state and Werner states \cite{E-thesis}},
but in spite of the werner states,  ${\textit{f}}^{\,\,q}$ can be a
positive number for Heisenberg chains. This means that, there exists
a channel such that it can teleport more entangled initial state
with more fidelity! but it is not a useful claim because, when we
choose the parameters of the channel such that
${\textit{f}}^{\,\,q}>0$ then ${\textit{f}}^{\,\,c}$ decreases and
ultimately $F(\rho _{in} ,\rho _{out})$ becomes smaller than
$\frac{2}{3}$ which means that  the entanglement teleportation of
mixed state is inferior to classical communication.
 Thus, to obtain the same  proper fidelity, the larger entangled channel
 are required for larger entangled initial state. The fig. 7  emphasize the above notes. \\
 The average fidelity $F_A$ is another useful concept for characterizing the quality of
 teleportation. The Average fidelity $F_A$ of teleportation can be obtained by
averaging $F(\rho_{in},\rho_{out})$ over all possible initial
states:
\begin{eqnarray} \label{FA1}
F_A  = \frac{{\int_0^{2\pi } {d\phi } \int_0^\pi
{F(\rho_{out},\rho_{in})\sin \theta d\theta } }}{{4\pi }}\,,
\end{eqnarray}
and for our model $F_A$ can be written as
\begin{eqnarray} \label{FA2}
F_A =\frac{{\xi ^2 \cosh ^2 \beta \eta + e^{2\beta J_z }(2\xi ^2
\cosh ^2 \beta \xi  + J^2 \sinh ^2 \beta \xi )}}{{3\xi ^2 (\cosh
\beta \eta  + e^{\beta J_z }\cosh \beta \xi )^2 }}.\,\,
\end{eqnarray}
In the case of an isotropic XXX Heisenberg chain in the absence of
magnetic field with spin-orbit interaction, this equation gives the
same as results of ref. \cite{Zh}. The function $F_A$ is dependent
on the channel's parameters. The fig. 8 gives a plot of $F_A$,
$C_{out}$, $F(\rho_{in}, \rho_{out})$ and $C_{channel}$ versus the
channel's parameters . This figure shows that, in the case of $J_z
>0$, $F_A$, $F(\rho_{in}, \rho_{out})$ and $C_{out}$ decrease when
$D$ increases. In this case $F_A$ approaches $\frac {2}{3}$ for
large values of $D$.

\begin{figure}
\epsfxsize=10cm \ \centerline{\hspace{0cm}\epsfbox{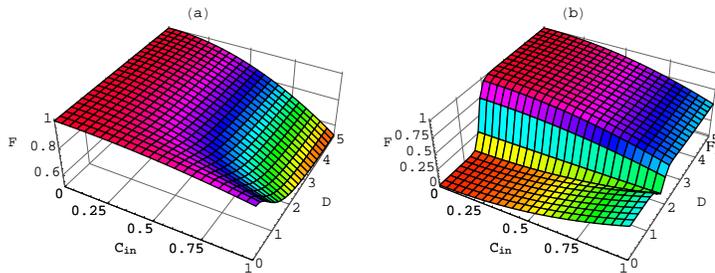}} \
\caption{(color online) $F(\rho_{in},\rho_{out})$ vs. $C_{in}$ and
$D$ for a) $J_z >0$ and b) $J_z <0$. Where $J=1, T=0.1, B=1, b=0.5$
and $\gamma=0.3$ }\label{f7}
\end{figure}

In contrast, in the case of $J_z <0$ and for small values of $D$,
$F_A$ and $F(\rho_{in}, \rho_{out})$ has a constant value (smaller
than $\frac {2}{3}$) and $C_{out}$ is zero. As $D$ becomes larger
than a threshold value, $C_{channel}$ undergoes a revival and then
decreases for larger values of $D$. Since in the revival region,
$C_{channel}$ has its maximum value, $F_A$, $F(\rho_{in},
\rho_{out})$ and $C_{out}$ increase in this region such that for
particular interval of $D$, $F_A$ becomes larger than $\frac {2}{3}$
and ultimately, tends to $\frac {2}{3}$ for larger $D$. In summary,
the fidelity of teleportation and entanglement of replica state are
dependent on the entanglement of channel which is tunable by the
channel's parameters (such as $D, b, B,...$). The effect of D is
more desirable in the case of $J_z <0$. In this case for a certain
values of $D$, $F_A$ becomes greater than $\frac{2}{3}$, this make
the channel useful for performance the teleportation protocol. For
large $D$, $F_A$ tends to $\frac{2}{3}$.

\begin{figure}
\epsfxsize=10cm \ \centerline{\hspace{0cm}\epsfbox{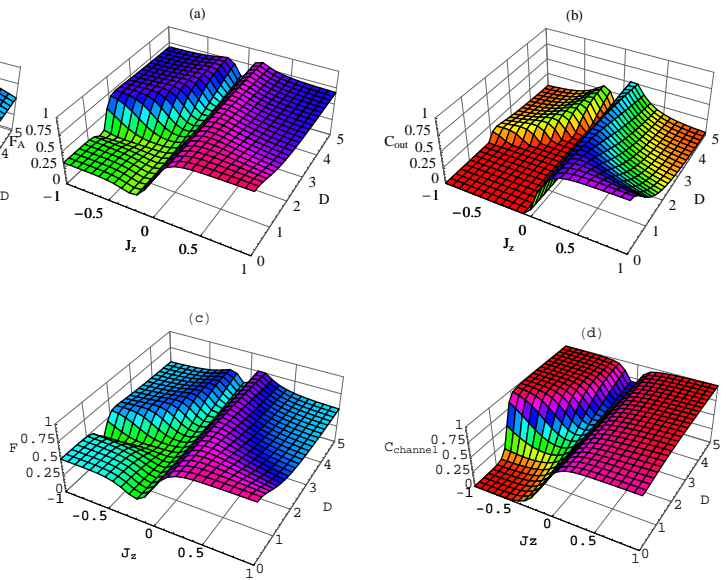}} \
\caption{(color online) a) Average fidelity (a), Output entanglement
(b), Fidelity between $\rho_{in}$ and $\rho_{out}$ (c) and the
entanglement (d) of channel versus $J_z$ and $D$. Where $J=1, T=0.1,
B=1, b=0.5, \gamma=0.3$ and $C_{in}=1$}\label{8}
\end{figure}

\section{Discussion}
The entanglement of a two-qubit XYZ Heisenberg system in
 the presence of an inhomogeneous magnetic field and spin-orbit interaction is investigated.
  By turning on the spin-orbit interaction we can change the behavior of the system
without manipulating the other parameters. We have shown that the
critical values of $B, \, b$ and $T$ are adjustable by $D$ and we
can improve the critical values of $B, \, b$ and T. Increasing $D$
cause to volume enhancement of revival region and also enhancement
of entanglement in the revival region. Also, entanglement
teleportation via two copies of above two-qubit system is studied.
We have shown that, by introducing SO interaction, the entanglement
of replica state and fidelity can be increased for the case of $J_z
<0$, in spite to the case $J_z>0$. When $D$ becomes very large, the
fidelity approaches $\frac {2}{3}$, which is the maximal value for
the classical communication.


\newpage
\begin{thebibliography}{99}
\bibitem {EPR} A. Einstein, B. Podolsky and N. Rosen, Phys. Rev.
47, 777 (1935)
\bibitem {S} E. Schr\"{o}dinger, Naturwiss. 23, 807 (1935)
\bibitem {NC-book} M. A. Neilsen and I. L. Chuang, \textit{Quantum computation and quantum information}
 (Cambridge University Press) (fifth printing 2004)
\bibitem {A-book} J. Audretsch, \textit{Entangled system}, WILEY-VCH Verlag (2007)
\bibitem {CR-2004} N. Canosa and R. Rossignoli, Phys. Rev. A 69, 052306 (2004)
\bibitem {W} S. Hill and W. K. Wooters, Phys. Rev. Lett. 78 (1997; W. K. Wootres, Phys. Rev. Letts., 80, 2245 (1998)
\bibitem {E-thesis} J. Eisert, \textit{Entanglement in Information Theory},
Ph.D thesis (Februray 2001)
\bibitem {N} M. A. Nielsen, e-print quant-ph/0011036
\bibitem {GKV} D. Gunlycke, V. M. Kendon, V. Vedral, Phys. Rev. A
64, 042302 (2001)
\bibitem {ABV} M. C. Arnesen, S. Bose, V. Vedral, Phys. Rev. Lett. 87, 017901 (2001)
\bibitem {KS} G. Lagmago Kamta and Anthony F. Starace, Phys. Rev. Lett., 88, 107901 (2002)
\bibitem {SCC} Y. Sun, Y. chen and H. Chen, Phys. Rev. A 68, 044301 (2003)
\bibitem {ZL} G. F. Zhang and S. Li, Phys. Rev. A 72, 034302 (2005)
\bibitem {AK} M. Asoudeh and V. Karimipour, Phys. Rev. A 71, 022308 (2005)
\bibitem {ZSGL} L. Zhou, H. S. Song, Y. Q. Guo and C. Li, Phys. Rev. A 68, 024301 (2003)
\bibitem {R} G. Rigolin, Int. J. Quant. Inf. 2, 393 (2004)
\bibitem {YGZS} G. H. Yang, W. B. Gao, L. Zhou and H.S. Song, quant-ph/0602051 (2006)
\bibitem {D} I. Dzyaloshinski, J. Phys. Chem. Solids 4, 241 (1958)
\bibitem {M1} T. Moriya, Phys. Rev. 117, 635 (1960)
\bibitem {M2} T. Moriya, Phys. Rev. Lett.4, 228 (1960)
\bibitem {M3} T. Moriya, Phys. Rev. 120, 91 (1960)
\bibitem {BS} N. E. Bonesteel and D. Stepaneko, Phys. Rev. Lett. 87,
207901 (2001)
\bibitem {BLs} G. Burkard and D. Loss, Phys. Rev. Lett. 88,
047903 (2002)
\bibitem {WL} L. A. Wu and D. Lidar, Phys. Rev. A 66,062314 (2002)
\bibitem {XW} X. Wang, Phys. Lett. A 281, 101 (2001)
\bibitem {LDi} D. Loss and D.P. Divincenzo, Phys. Rev. A 57, 120
(1998)
\bibitem {DD} D. DiVincenzo, Phys. Rev. A 51,1015 (1995)
\bibitem {CL} W. A. Coish and D. Loss, e-print cond-mat/0606550
(1998) and G. Burkard, D. Loss and D. P. Divincenzo , Phys. Rev. B
59, 2070 (2002)
\bibitem {X-thesis} H. Xiong, \textit{Coherent-Induced Entanglement}, Ph.D
thesis (May 2006)
\bibitem {CFH} D. Cory, A. Fahmy and T. Havel, Proc. Natl. Acad.
Sci. USA 94,1634 (1997)
\bibitem {CGK} I. L. Chuang, N. A. Gershenfeld and M. Kubinec, Phys.
Rev. Lett. 80,3408 (1998)
\bibitem {JMH} J. A. Jones, M. Mosca and R. H. Hansen, Nature (London)
393, 344 (1998)
\bibitem {PEL} A. Pfund, K. Ensslin and R. Leturcq, Phys. Rev. B 76, 161308(R) (2007)
\bibitem {ZZZS} N. Zhao, L. Zhong, J. L. Zhu and C. P. Sun, Phys. Rev. A 74, 075307 (2006)
\bibitem {DE} S. Debald and C. Emary, Phys. Rev. Lett. 94, 226803
(2005)
\bibitem {GBL} V. N. Golovach, M. Borhani, and D. Loss, Phys. Rev. B 74, 165319
(2006); D. V. Bulaev and D. Loss, Phys. Rev. Lett. 98, 097202 (2007)
\bibitem {FSF} C. Flindt, A. S. Srensen, and K. Flensberg, Phys. Rev. Lett. 97, 240501 (2006)
\bibitem {BBCJPW} C. H. Bennett, G. Brassard, C. Crepeau, R. Josza, A. Peres and W. K. Wooters,
 Phys. Rev. Letts. 70, 1895-1899 (1993)
\bibitem {Pop} S. Popescu, Phys. Rev. Letts. 72, 797-799 (1994)
\bibitem {HHH} M. Horodecki, P. Horodecki and R. Horodecki, Phys.
Rev A 60, 1888 (1999)
\bibitem {BB} G. Bowen and S. Bose, Phys. Rev. Lett. 87, 2679011
\bibitem {LK} J. Lee, M. S. Kim, Phys. Rev. Letts. 84 ,4236 (2000)
\bibitem {Y} Y. Yeo, Phys. Rev. A 66, 062312 (2002) and Y. Yeo, quant-ph/023014 (2002)
\bibitem {YLLY} Y. Yeo, T. Liu, Y. Lu and Q. Yang, J. Phys. A: Math.
Gen. 38, 3235 (2005)
\bibitem {Zh} G. F. Zhang, Phys. Rev. A 75, 034304 (2007)
\bibitem {S-thesis} S. Olav Skr{\o}vseth, e-print quant-ph/0612133
(2006)
\bibitem {S-book} S. Sachdiv, \textit{Quantum phase transition}
(Cambridge University Press, Cambridge, UK,1999)
\bibitem {RJ} R. Josza, J. mod. Opt. 41, 2315, (1994)
\end{thebibliography}
\end{document}